\documentclass[useAMS,usenatbib]{mn2e}

\usepackage{epsfig}
\usepackage{longtable}
\usepackage{times}

\usepackage{amsmath}
\usepackage{amssymb}
\def \inte {$INTEGRAL$}
\def \xmm {$XMM$-$Newton$}

\def \chandra {$Chandra$}

\def \src {IGR~J19140+0951}
\def \igr {IGR~J19140+0951}

\def \hcm {\hbox {\ifmmode $ atom cm$^{-2}\else atom cm$^{-2}$\fi}}
\def \arcmin {\hbox{$^\prime$}}
\def \arcsec {\hbox{$^{\prime\prime}$}}

\def \apj {ApJ}

\def \apjl {ApJL}
\def \apjs {ApJS}
\def \aap {A\&A}
\def \aaps {A\&AS}

\def \mnras {MNRAS}
\def \nat {Nature}

\def \iaucirc {IAU Circ.}

\def\xmm {\emph{XMM-Newton}}
\def\cxo {\emph{Chandra}}

\newcommand{\be}{\begin{equation}}

\newcommand{\ee}{\end{equation}}

\title[QPOs in \igr]{\xmm\ discovery of  mHz quasi-periodic oscillations in the high mass X-ray binary \src
}
\author[Sidoli et al.]{L.~Sidoli,$^{1}$\thanks{E-mail: sidoli@iasf-milano.inaf.it} P.~Esposito,$^{1,2}$ S.~E.~Motta,$^{3}$  G.~L.~Israel,$^{4}$ and G.~A.~Rodr\'iguez Castillo,$^{4}$ \\
$^{1}$INAF, Istituto di Astrofisica Spaziale e Fisica Cosmica, Via E.\ Bassini 15,   I-20133 Milano,  Italy   \\
$^{2}$Anton Pannekoek Institute for Astronomy, University of Amsterdam, Postbus 94249, NL-1090-GE Amsterdam, The Netherlands \\
$^{3}$Department of Physics, Astrophysics, Denys Wilkinson Building, University of Oxford, Keble Road, Oxford OX1 3RH, UK \\
$^{4}$INAF, Osservatorio Astronomico di Roma, Via Frascati 33, I-00040 Monteporzio Catone, Italy \\
}

\begin{document}

\date{Accepted   Received}

\pagerange{\pageref{firstpage}--\pageref{lastpage}} \pubyear{2015}

\maketitle

\label{firstpage}

\begin{abstract}
We report  on the discovery of mHz quasi-periodic oscillations (QPOs) from the high mass X-ray binary (HMXB) \igr, 
during a 40 ks \xmm\ observation performed in 2015, which caught the source in its faintest state ever observed. 
At the start of the observation, \src\ was at a low flux of  2$\times$10$^{-12}$~erg~cm$^{-2}$~s$^{-1}$ 
(2-10 keV; L$_{\rm X}$=3$\times$10$^{33}$~erg~s$^{-1}$ at 3.6 kpc), then its emission rised reaching 
a flux $\sim$10 times higher, in a flare-like activity.
The investigation of the power spectrum reveals the presence of QPOs, detected only in the second part of the observation, with a strong peak  
at a frequency of 1.46$\pm{0.07}$~mHz, together with higher harmonics.
The X--ray spectrum is highly absorbed  (N$_{\rm H}$=$10^{23}$~cm$^{-2}$), well fitted by 
a power-law with a photon index  in the range 1.2-1.8.
The re-analysis of a $Chandra$ archival observation shows a modulation at  $\sim$0.17$\pm{0.05}$~mHz, very likely 
the neutron star spin period (although a QPO cannot be excluded).
We discuss the origin of the 1.46~mHz QPO in the framework of both disc-fed and wind-fed HMXBs, 
favouring the quasi-spherical accretion scenario. 
The low flux observed by XMM-Newton leads to about three orders of magnitude the source dynamic range, overlapping with
the one observed from Supergiant Fast X--ray Transients (SFXTs). However, since its duty cycle is not as low as in SFXTs, \src\ is 
an intermediate system between persistent supergiant HMXBs and SFXTs, suggesting a smooth transition between these two sub-classes.
\end{abstract}


        \section{Introduction\label{intro}}

\igr\  (formerly known as IGR J19140+098) was discovered  in 2003 March during an \inte\
observation targeted on the microquasar GRS~1915+105 (\citealt{Hannikainen2003, Hannikainen2004}).
A follow-up observation with $RossiXTE$ PCA detected the source at 10$^{-10}$~erg~cm$^{-2}$~s$^{-1}$ (2--10~keV) 
and 2$\times$10$^{-10}$~erg~cm$^{-2}$~s$^{-1}$ (10--60 keV). The source was variable on timescales longer than 100~s,
with a power-law spectrum (N$_{\rm H}$=6$\times10^{22}$~cm$^{-2}$, photon index, $\Gamma$, of 1.6) 
and an iron line complex (equivalent width EW=550~eV; \citealt{Swank2003}). 
The  flux dynamic range of 150 in the energy band 1-20 keV \citep{Rodriguez2006}, 
the iron line  and the absorbing column density properties were interpreted as
indicative of an HMXB nature \citep{Rodriguez2005, Hannikainen2004, Prat2008}. 
This suggestion was 
strengthened by the discovery of a periodicity of 13.558(4)~days (epoch of maximum flux on MJD 51593.4$\pm{0.32}$) in  the 
$RossiXTE$ ASM light curve, likely the orbital period of the system \citep{Corbet2004}.
A refined period of 13.552(3)~days was later determined by \citet{Wen2006}, again analysing $RossiXTE$ ASM data.

The confirmation that \igr\ is indeed an HMXB came from the identification of the optical counterpart, the star 2MASS~19140422+0952577,
which was possible thanks to the $Chandra$ sub-arcsec X--ray position \citep{zand2006}.
Several authors investigated its optical/infrared properties (\citealt{zand2006}, \citealt{Nespoli2007}, \citealt{Hannikainen2007}, 
\citealt{Chaty2008}, \citealt{Rahoui2008},  \citealt{Nespoli2008}, \citealt{Torrejon2010}), all confirming an early-type supergiant companion.
The most recent optical photometry performed by \citet{Torrejon2010} indicates a B0.5~Ia star located at a distance of $\sim$3.6~kpc.

The possible identification with the  X--ray source EXO~1912+097 (\citealt{Lu1996}, \citealt{Lu1997}),  
suggested by several authors (\citealt{Hannikainen2004}, \citealt{Hannikainen2007}, \citealt{zand2004, zand2006}) 
appears rather uncertain (see Section~\ref{ls:skypos} for a discussion of the available literature). 
However, $RossiXTE$ ASM \citep{Corbet2004} and $BeppoSAX$ WFC \citep{zand2004} 
detected it at earlier times, before the \inte\ discovery. 

Since the absorbing column density is an order of magnitude higher than the interstellar one \citep{Nespoli2008, Prat2008},
\igr\ was suggested to belong to the so-called ``highly obscured HMXBs'' unveiled by \inte\ satellite \citep{Kuulkers2005}.

In this {\em paper}, we report on the discovery of mHz quasi-periodic oscillations (QPOs) in an \xmm\ observation performed in 2015 October,
together with a re-analysis of $Chandra$ archival data.

 	 \section{Observations and Data Reduction}
         \label{data_redu}

\subsection{\cxo}
\label{chandra_red}

A \cxo\ observation of \src\ was carried out on 2004 May 11 (obs. ID 4590) for a 
total exposure time of 20.1~ks \citep{zand2006}. 
The source was positioned in the back-illuminated ACIS-S3 CCD and the instrument was operated 
in full-imaging timed-exposure mode (with no gratings), with a frame time of 3.24~s. 
We generated new level 2 event files the \cxo\ Interactive Analysis of Observations (\textsc{ciao}) software 
version 4.7 and following standard procedures. 
Since our results for the spectral analysis are essentially consistent with those 
of \citet{zand2006}, to which we refer the reader, in this paper we give only some 
details on the timing analysis that led to the discovery of the modulation of \src\ (Section \ref{pulsation}).

\subsection{\xmm}

The \xmm\ $Observatory$ carries three 1500~cm$^2$ X--ray
telescopes, each with an European Photon Imaging Camera (EPIC) detector
at the focus. Two of the EPIC detectors use MOS CCDs \citep{Turner2001} and one uses pn CCDs
\citep{Struder2001}. Reflection Grating Spectrometer (RGS; 0.4-2.1 keV)
arrays \citep{DenHerder2001} are located 
in the telescopes with MOS cameras at their primary focus. 

The source position was observed by \xmm\ on 2015,  October 26-27, for a net exposure of 40.2~ks (Obs. ID 0761690301).
The EPIC pn camera was operated in full frame mode, while both the MOS cameras were in small window mode.
The data were reprocessed using version 14 of the Science Analysis Software ({\sc sas}) with standard procedures.
Given the high absorption in the source direction which severely affected the observed emission up to 2~keV, 
the RGS data led to insufficient counts to afford a meaningful spectroscopy, so we will not discuss them further.

\begin{figure}
\centering
\resizebox{\hsize}{!}{\includegraphics[angle=0]{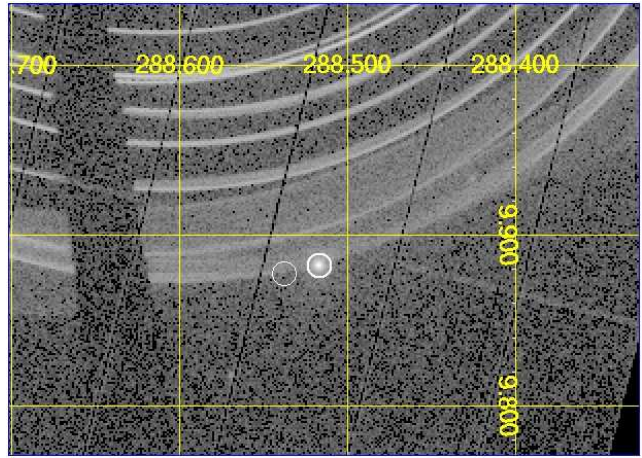}}
\caption{Close-up view of the central part of the EPIC pn image (2-12 keV). Sky coordinates are R.A. and Dec (J2000). Two white circles mark the 
extraction regions for the background (thin circle) and the source counts (thick circle). Stray light contamination from a bright source outside the FOV is evident.
}
\label{lsfig:ima}
\end{figure}

The EPIC field of view (FOV) was contaminated by the stray light coming from a bright source
outside the FOV (Fig.~\ref{lsfig:ima}), very likely the microquasar GRS~1915+105.

For the spectral analysis, we used  EPIC data.
We extracted source counts from circular regions with radii of
25\arcsec\ radius (EPIC pn and MOS2) and 30\arcsec\ radius (MOS1), 
using PATTERN from 0 to 4 (mono- and bi- pixel events) in the pn, 
and from 0 to 12 (up to 4-pixel events) in both MOS cameras.
Background counts were obtained from similar circular regions, offset from the source position.
Since the pn operated in full frame mode, it was possible to 
optimize the choice for the background region nearby the source, to subtract better the stray light contamination (see Fig.~\ref{lsfig:ima}).
Background regions from different CCDs, farther away from the source, were selected in both MOS cameras, 
which were operated in small window mode. 
However, the choice of the background spectra seemed not to affect significantly the spectral results: we obtained similar spectral parameters
fitting only pn spectrum compared to analysis of the joint pn and  MOS spectra together.

The background level was stable along the observation, so no further filtering was applied.
Appropriate response matrices were generated
using the {\sc sas}  tasks {\sc arfgen} and {\sc rmfgen}.
All spectral uncertainties  are given at 90\% confidence level for
one interesting parameter. 
Spectra were grouped so to have a minimum of 25 counts per bin.
We fitted the three EPIC spectra simultaneously, adopting normalization factors to account for uncertainties in instrumental responses.
Fluxes are quoted adopting the pn camera response matrix. 
In the spectral fitting we adopted the interstellar abundances of \citet{Wilms2000}  
and  photoelectric absorption cross-sections of \citet{bcmc}, using the absorption model  {\sc TBabs} in {\sc xspec}.

  	\section{Analysis and Results\label{result}}

\subsection{Timing analysis of \cxo\ data}
\label{pulsation}

\begin{figure*}
\centering
\resizebox{\hsize}{!}{\includegraphics[angle=-90]{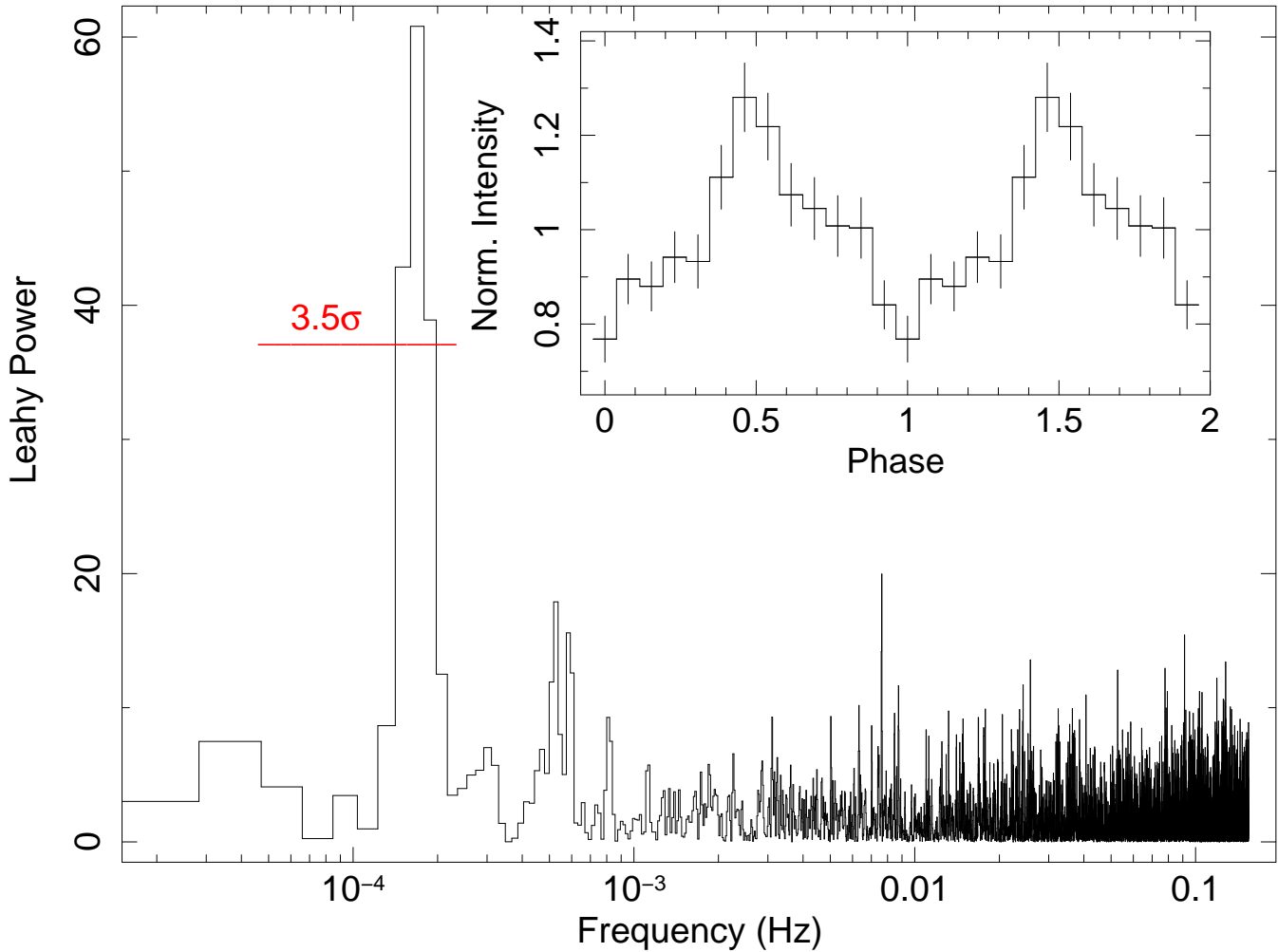}}
\caption{\label{powerspec} Power spectrum of \src\ computed from the \cxo/ACIS data of 2004 May 11 (0.5--10~keV). 
The red horizontal line indicates the 3.5$\sigma$ significance level, which is essentially constant at all frequencies.
The inset shows the light curve folded at the period we measured ($P=5\,937$~s).
}
\end{figure*}


In the 2004 \cxo\ observation of \src, \citet{zand2006} noticed a possible 
modulation with a period of about 6.5~ks. 
Indeed, a signal at a similar period ($\sim$5.9~ks) was discovered 
by the detection algorithm of the \emph{Chandra ACIS Timing Survey} (CATS) 
project\footnote{The CATS project is a Fourier-transform-based systematic and 
automatic search for new pulsating sources in the \cxo\ ACIS public archive. 
So far more than half a million light curves were analyzed and the effort 
yielded dozens of previously unknown X-ray pulsators.} during an authomatic 
search of the same data set (Israel et al., submitted; see also \citealt{Esposito2013, Esposito2015}).

Fig.\,\ref{powerspec} shows the Fourier periodogram that led to the discovery of 
the signal in \src,  normalized according to \citet{Leahy1983}. The number of searched frequencies is 8192. 
A peak at frequency $\nu \simeq 0.168$~mHz ($P=1/\nu\simeq5\,950$~s) stands above the 3.5$\sigma$ significance level, 
which was estimated taking into account the number of frequencies searched and the noise components of 
the spectrum following \citet{Israel1996}. 
Given the small number of modulation cycles sampled by the \cxo\ observation (roughly 3.5), 
to measure the period we simply  fitted a sinusoid curve to the \cxo\ light curve. 
This returned the value $P = 5\,937\pm219$~s (90\% confidence level).  
The inset of Fig.\,\ref{powerspec} shows the light curve folded at this period.
A pulsed fraction (defined as the semi-amplitude of the sinusoid divided by the average count rate) 
of 20$\pm$3\% was inferred from the fit. 

The quality factor of the signal (defined as $Q = \nu/\Delta\nu$, the
ratio of the frequency $\nu$ of the peak to its full width at
half-maximum {\bf $\Delta\nu$}) $Q\simeq3$ is comparable to the typical
values of QPOs. However, we notice that the low Fourier resolution of
our data  (see Fig.\,\ref{powerspec}) would not allow us
to distinguish the case of a QPO from that of a strictly coherent
pulsation  around 0.1~mHz.
Furthermore, see the discussion in Sect.~\ref{sect:qpo} supporting the interpretation of this signal
as the NS spin period.
The source flux during the  \cxo\ observation was around 10$^{-10}$~erg~cm$^{-2}$~s$^{-1}$ \citep{zand2006}, translating into an X-ray luminosity of $\sim$10$^{35}$~erg~s$^{-1}$ at 3.6 kpc.

\subsection{Timing analysis of \xmm\ data}
\label{xmm_qpo}

\begin{figure}
\begin{center}
\includegraphics[height=8.7cm,angle=-90]{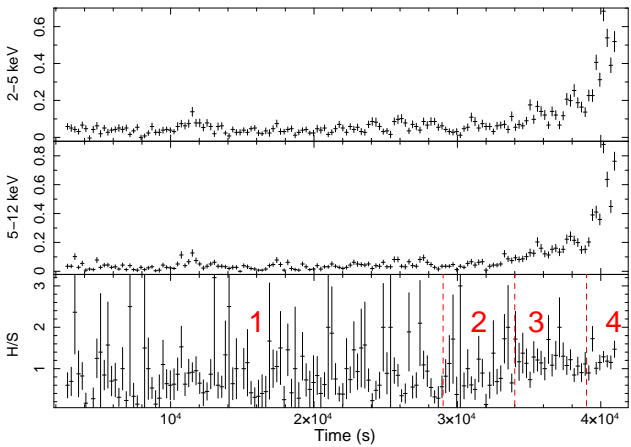}
\end{center}
\caption{\igr\ EPIC pn background subtracted X--ray light curves (below and above 5 keV; bin time 256~s), together with their 
hardness ratio. Numbers in the lower panel mark the four time intervals chosen for the temporal selected spectroscopy. 
}
\label{lsfig:2lc}
\end{figure}

The \src\ EPIC pn background-subtracted light curve in two energy ranges is shown in Fig.~\ref{lsfig:2lc}.
It starts with a low, although quite variable, emission. Then a flare-like behaviour is present in the second half of the observation.
Folding the whole light curve at the periodicity found in the $Chandra$ observation, a modulation is evident in the data.
However,  it does not provide conclusive proof that the modulation found in \cxo\ data is present also in the \xmm\ data, 
since folding EPIC light curve at diffent values of the putative period, results in apparent modulation as well.
Moreover, \xmm\ data are not constraining, as the upper limit on the  pulsed fraction is $>$100\%.

\begin{figure*}
\centering
\resizebox{\hsize}{!}{\includegraphics[angle=0]{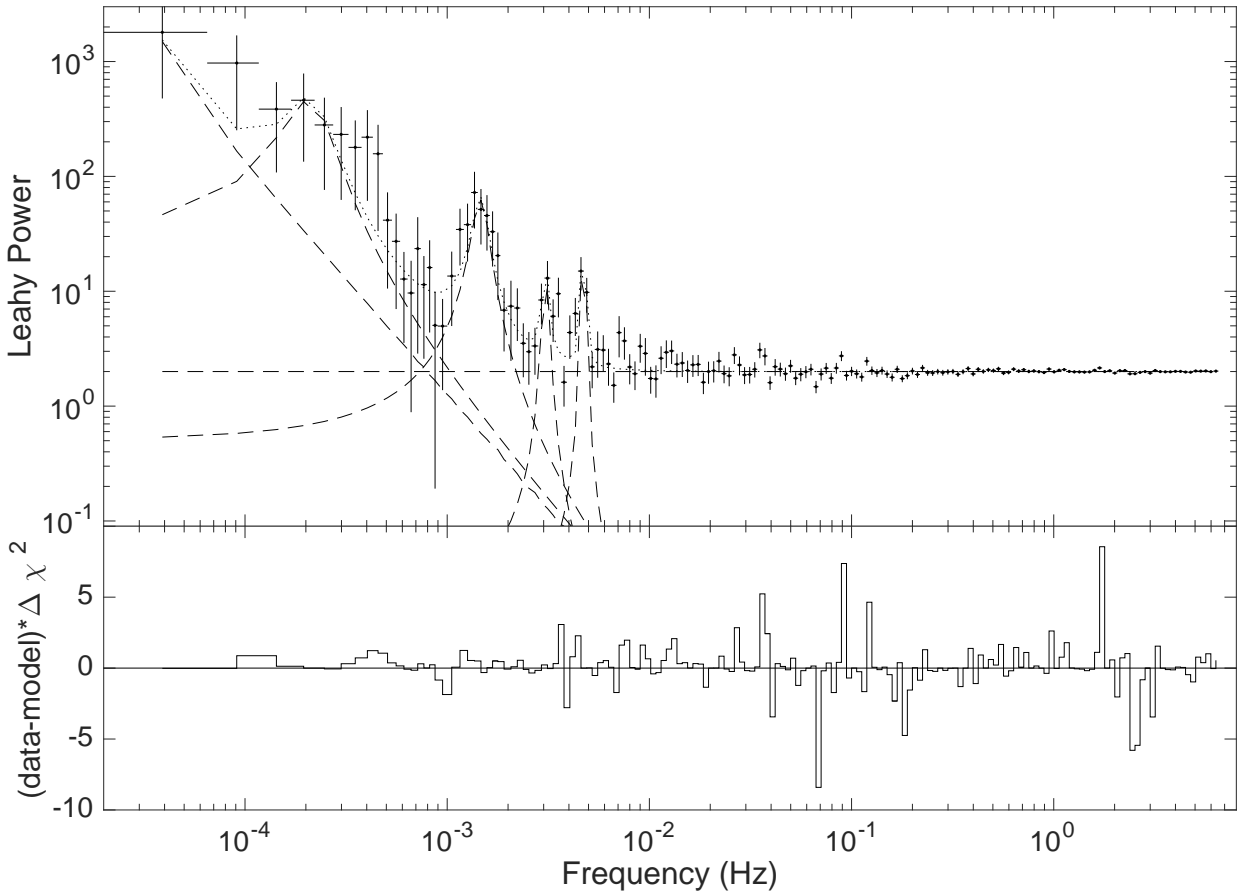}}
\caption{\label{qpo} Power spectrum of \src\ computed from the  whole \xmm\ EPIC pn data of 2015 (0.1--10~keV). 
}
\end{figure*}


We performed the timing analysis of \src\ computing the power density
spectra (PDS) from the EPIC pn data using custom software under IDL  in the full
EPIC pn energy band (0.1-10~keV). We produced one single PDS for the entire
dataset (Fig.~\ref{qpo}), between ~0.03~mHz (our frequency resolution)  and the maximum possible
Nyquist frequency ($\sim$6.8~Hz, set by the time resolution of our observation,
$\sim$0.073~s), normalized according to \citet{Leahy1983} and converted to square
fractional rms \citep{Belloni1990}. We did not subtract the contribution due to
Poissonian noise, but we fitted it afterwards. 

The total fractional variability of the PDS is surprisingly high, with almost
93~\% of the emission variable. The PDS shows a broad-band, possibly flat-top
noise component and a clear fairly narrow feature  at $\sim$1.5~mHz, with
two other narrow features, apparently harmonically related to the first one.
We carried out the PDS fitting with the {\sc xspec} fitting package by using
a one-to-one energy-frequency conversion and a unit response. 
Following \citet{Belloni2002}, we fitted the noise components 
with a suitable number of broad Lorentzians (in this case one zero-centred
broad Lorentzian and a narrower one at slightly higher frequencies), while we fitted the QPO peaks with
narrow Lorentzians. Even though the three QPO peaks seem to be harmonically related, we did not link
the peak frequencies in the fit.
A constant component was added to the model to take into account the contribution
of the Poissonian noise.
Our fit shows that the strongest peak is highly significant (4$\sigma$) and is
centered at 1.46$\pm{0.07}$~mHz, with a FWHM of 0.25~mHz $\pm{ 0.07}$~Hz and amplitudes
42($\pm{5}$)\% rms. The second and third harmonics are detected with a 2.3 and 3.4
$\sigma$ significance, respectively, at  frequencies consistent with being harmonics of the strongest peak.
The QPO and its harmonics are only detected in the second part of our
observation, in correspondence of the large flare starting at about 30~ks from
the observation start, while only broad band noise and Poisson noise are visible
in the first half of the observation. Since the QPO is detected at very low
frequency, the total exposure of our observation limits our analysis and
prevents us from a further investigation of the transient nature of this feature.
We also notice that the QPO  shows a dependence on energy, being clearly
present only above 3~keV. 

Since our observation is affected by stray lights caused by the presence of a
close-by bright source out of the FOV  (probably the bright black hole (BH)
binary GRS~1915+105), we performed some additional timing analysis on the
background emission as well, in order to exclude possible contamination from 
GRS~1915+105 or other sources in the field. Following the steps outlined above, we
computed PDS from the background emission, both close to the source and on a CCD
region that is highly contaminated by the stray lights emission. 
In both cases, we did not detect any significant variability from the background, neither as a broadband noise nor 
in the form of a narrow feature.
Therefore, we can reasonably exclude that the QPO is of instrumental origin or due to
contamination from another, variable, source. 

\subsection{\xmm\ spectroscopy}
\label{sec:epic_spec}

\begin{figure*}
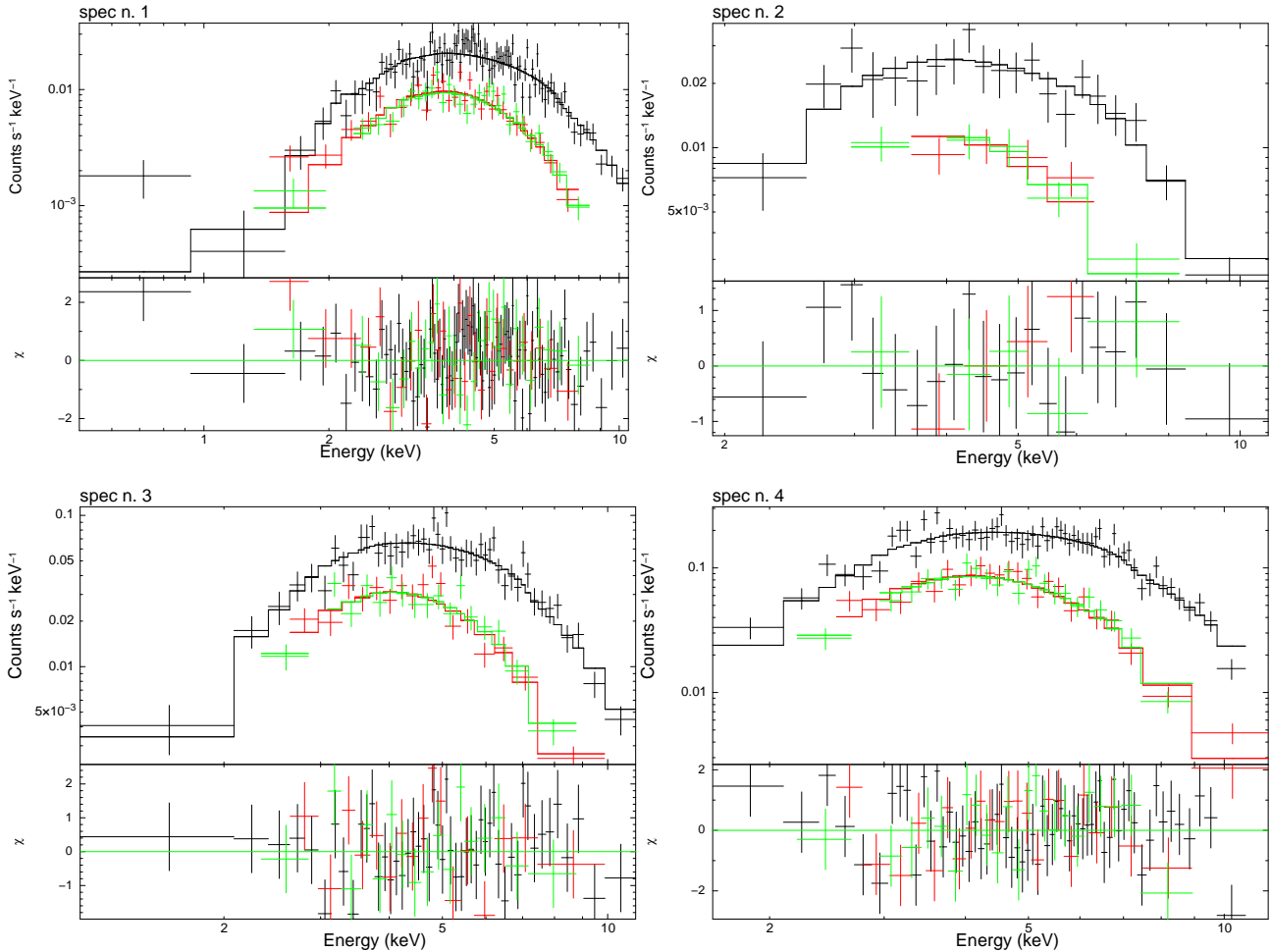

\begin{tabular}{cccc}
\includegraphics[height=8.5cm, angle=-90]{./fig5a.ps} 
\includegraphics[height=8.5cm, angle=-90]{./fig5b.ps} \\
\includegraphics[height=8.5cm, angle=-90]{./fig5c.ps} 
\includegraphics[height=8.5cm, angle=-90]{./fig5d.ps} \\
\end{tabular}
\caption{Spectral results from the temporal selected analysis. 
The four spectra correspond to the results reported in Table~\ref{lstab:spec} for the
power-law model. From left to right, top to bottom, count spectra number 1, 2, 3 and 4 are shown, together with their residuals 
in units of standard deviations from the power-law best-fit.  
In each panel, the upper spectrum (in black) marks EPIC pn, while lower spectra (in red and green) indicate MOS1 and MOS2, respectively.
}
\label{spec}
\end{figure*}

Two background subtracted EPIC pn light curves of \igr\ in two energy ranges  are shown in
Fig.~\ref{lsfig:2lc}, together with their hardness ratio.
A clear hardening is present during the last half of the exposure, where the source intensity is increasing.
Given the hardness ratio variability along the EPIC exposure, we extracted 4 time-selected 
spectra (for each camera) from 4 time intervals, as shown in Fig.~\ref{lsfig:2lc}.
A joint fit of pn, MOS1 and MOS2 spectra  
with an absorbed power-law model  (see Fig.~\ref{spec}) or adopting  a simple absorbed black body, 
resulted in the parameters reported in Table~\ref{lstab:spec}.
In the same Table (last column) we also report  the spectroscopy of the time-averaged spectrum extracted from the whole 40~ks exposure.

\begin{figure*}
\begin{center}
\includegraphics[height=13.50cm,angle=0]{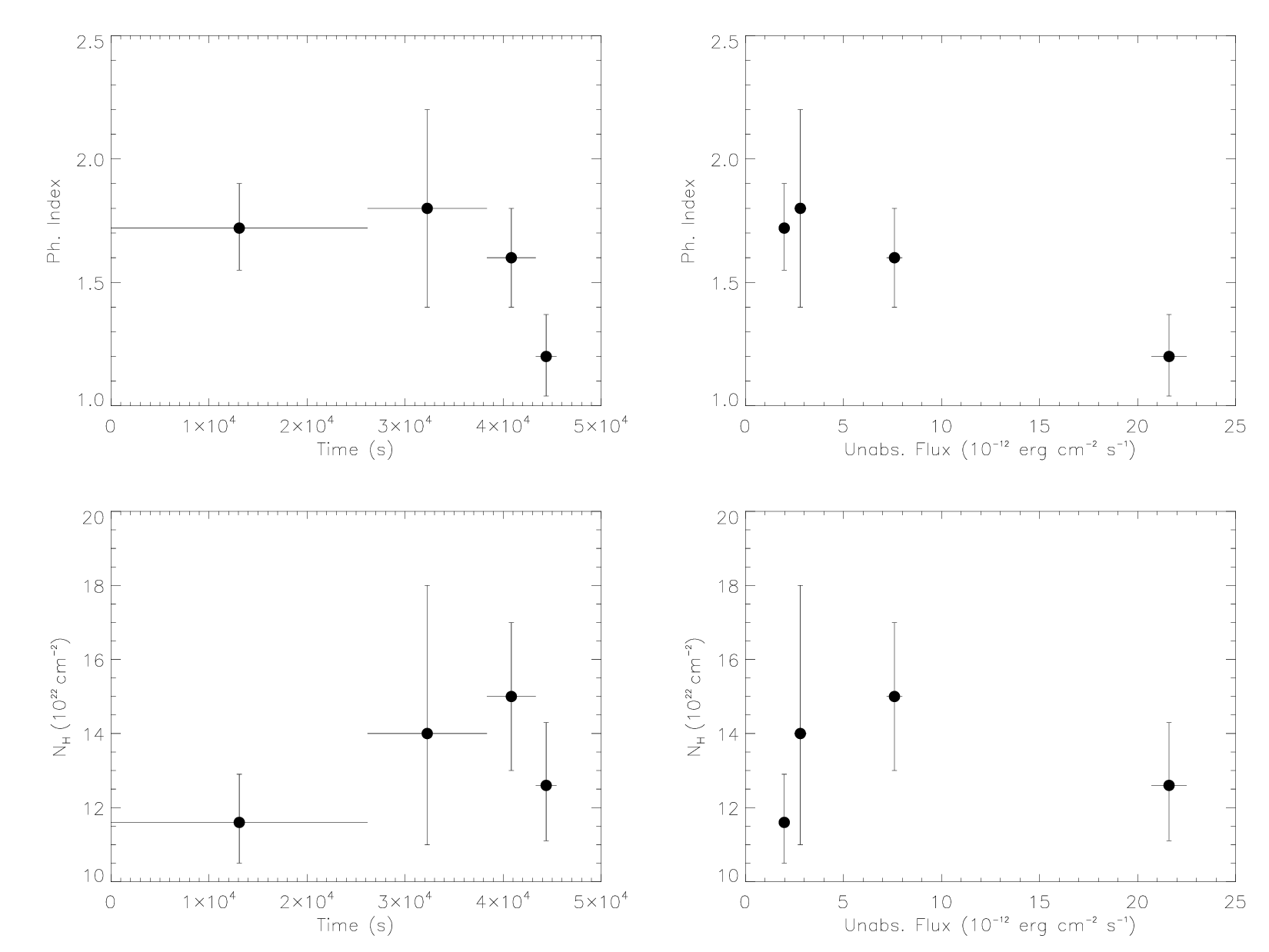}
\end{center}
\caption{EPIC time-resolved spectroscopy using an absorbed power-law model (Table~\ref{lstab:spec}).
}
\label{lsfig:pow}
\end{figure*}

\begin{figure*}
\begin{center}
\includegraphics[height=13.50cm,angle=0]{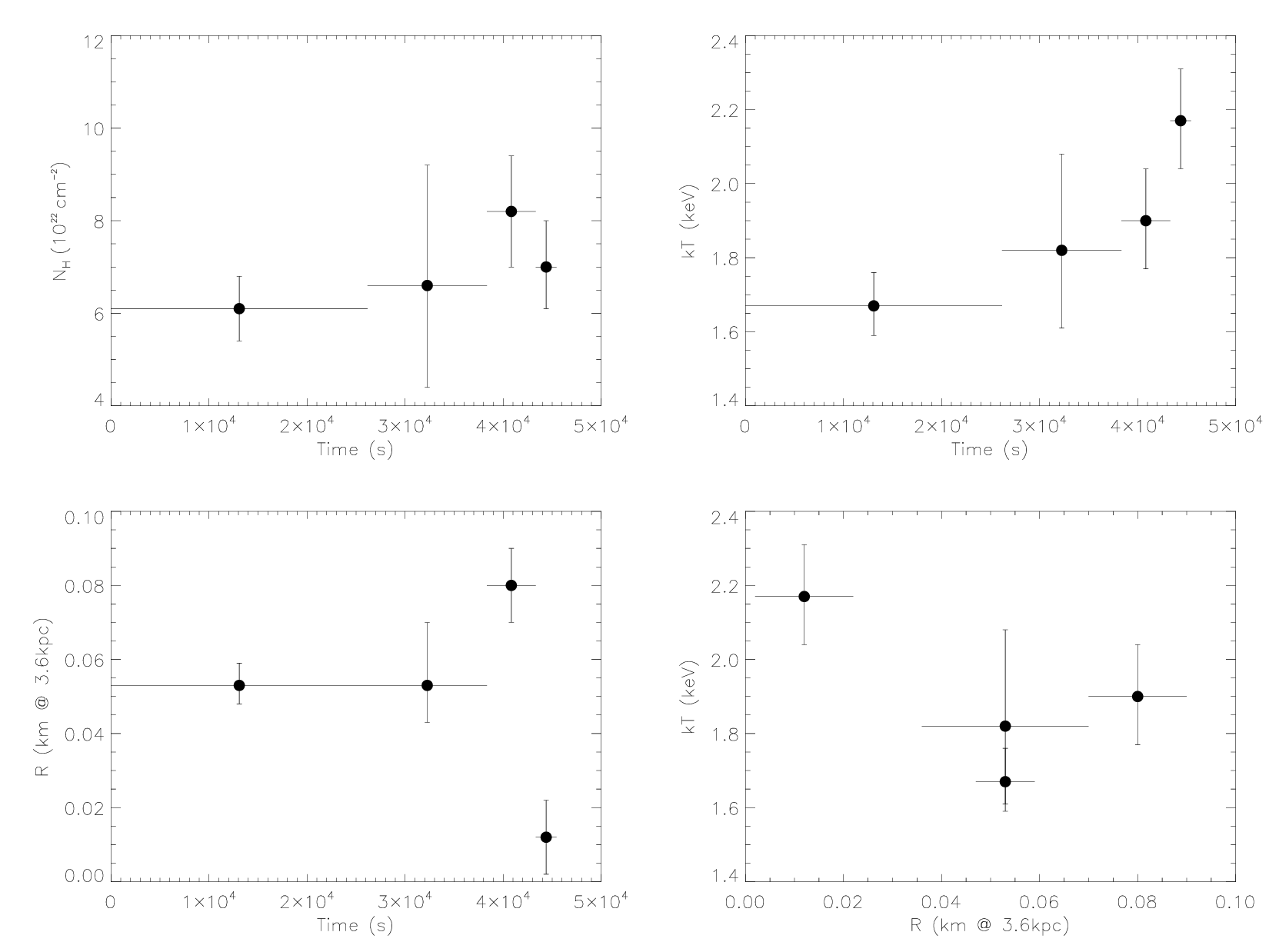}
\end{center}
\caption{EPIC time-resolved spectroscopy using an absorbed black body model (Table~\ref{lstab:spec}).
}
\label{lsfig:bb}
\end{figure*}

Besides the flux, there is evidence for a variability in the hardness (power-law photon index or black body temperature),
while the absorbing column density remains constant within the uncertainties (see Figs.~\ref{lsfig:pow} and ~\ref{lsfig:bb}).
Therefore, we next fixed the absorbing column density to the value obtained in the time-averaged spectroscopy,
and re-fitted the joint pn, MOS1, MOS2 spectra with a power-law or a black body. The results obtained with a fixed column density
are reported in the same table. 
The spectral hardening with increasing flux near the end of the observation is even more evident in this case.

\begin{table*}
\begin{center}
\caption[]{Results of the  temporal selected (EPIC pn, MOS~1 and MOS~2) spectroscopy with two simple models, 
an absorbed power-law ({\sc pegpwrlw} in {\sc xspec}) and an absorbed black body model.
Numbers identify the temporal selected spectra as in Fig.~\ref{lsfig:2lc}, while the last column reports on the time-averaged spectroscopy.
Flux is in the 2--10~keV energy range
in units of 10$^{-12}$~erg~cm$^{-2}$~s$^{-1}$ and is corrected for the  absorption, 
L$_{2-10 keV}$ is the X-ray luminosity in units of 10$^{33}$~erg~s$^{-1}$ (assuming a 3.6~kpc distance), 
N$_{\rm H}$ (in units of $10^{22}$~cm$^{-2}$; {\sc TBabs} model in  {\sc xspec} with Wilms et al. (2000) abundances). 
Black body temperature, kT$_{\rm bb}$, is in keV. Black body radius, R$_{\rm bb}$, is in km at 3.6 kpc. The uncertainties on L$_{2-10 keV}$ include only the 90\% error on the unabsorbed flux.}
\begin{tabular}{llllll}
\hline
\noalign {\smallskip}
Power-law param   &           1                          &     2                         &    3                         &      4                        &            Time-averaged    \\
\hline
\noalign {\smallskip}
N$_{\rm H}$             &  $ 11.6 ^{+1.3} _{-1.1}$       &    $14 ^{+4} _{-3}$           &    $15    ^{+2} _{-2}$      &     $ 12.6 ^{+1.7} _{-1.5}$    &       $ 12.8 ^{+0.8} _{-0.7}$     \\
$\Gamma$                &  $1.72^{+0.18} _{-0.17}$       &    $1.8  ^{+0.4} _{-0.4}$     &    $1.6  ^{+0.2} _{-0.2}$   &     $1.20 ^{+0.17} _{-0.16}$   &       $1.56 ^{+0.09} _{-0.09}$    \\

Unabs. Flux             &    $1.98 \pm{0.08}$            &    $2.8 \pm{0.3}$             &         $7.6 \pm{0.4}$         &   $21.6 \pm{0.9}$           &        $3.89 \pm{0.09}$           \\
L$_{2-10 keV}$          &   $3.1 \pm{0.1}$               &    $4.3 \pm{0.5}$             &    $11.8 \pm{0.6}$             &    $33.5 \pm{1.4}$          &      $6.0\pm{0.1}$               \\
$\chi^{2}_{\nu}$/dof    &    1.088/146                   &     0.649/25                  &      0.983/79                  &      1.013/104              &       1.126/292                   \\
\hline
\noalign {\smallskip}
Power-law param   &           1                          &     2                         &    3                         &      4                        &           \\
\hline
\noalign {\smallskip}
N$_{\rm H}$             &  $ 12.8$ fixed                 &   $ 12.8$ fixed               &    $ 12.8$ fixed             &       $ 12.8$ fixed          &                      \\
$\Gamma$                &  $1.86^{+0.09} _{-0.09}$       &    $1.66  ^{+0.21} _{-0.21}$  &    $1.37  ^{+0.10} _{-0.10}$ &     $1.22 ^{+0.09} _{-0.09}$ &                       \\

Unabs. Flux             &    $2.11 \pm{0.05}$            &    $2.64\pm{0.14}$            &         $6.9 \pm{0.2}$         &   $21.7 \pm{0.5}$           &                        \\
$\chi^{2}_{\nu}$/dof    &    1.097/147                   &     0.631/26                  &      1.024/80                  &      1.004/105              &                      \\
\hline
BB param               &           1                     &     2                         &    3                           &      4                      &            Time-averaged    \\
\hline
N$_{\rm H}$             &  $ 6.1 ^{+0.7} _{-0.7}$         &    $6.6  ^{+2.6} _{-2.2}$     &    $8.2  ^{+1.3} _{-1.2}$      &     $ 7.0 ^{+1.0} _{-0.9}$  &        $ 6.8 ^{+0.5} _{-0.4}$      \\
kT$_{\rm bb}$           &  $1.67 ^{+0.09} _{-0.08}$      &    $1.82  ^{+0.26} _{-0.21}$  &    $1.90  ^{+0.14} _{-0.13}$   &     $2.17 ^{+0.14} _{-0.13}$ &     $1.89 ^{+0.06} _{-0.06}$       \\
R$_{\rm bb}$           &    $0.053 ^{+0.006} _{-0.005}$  &  $0.053 ^{+0.017} _{-0.010}$  &    $0.080  ^{+0.010} _{-0.010}$ & $0.120  ^{+0.010} _{-0.010}$ &    $0.061  ^{+0.004} _{-0.003}$    \\
$\chi^{2}_{\nu}$/dof    &   1.075/146                     &   0.655/25                    &    0.992/79                   &      0.949/104               &       1.113/292                  \\
\hline
BB param               &           1                     &     2                         &    3                           &      4                      &          \\
\hline
N$_{\rm H}$             &  $6.8$  fixed                  &     $6.8$  fixed              &     $6.8$  fixed               &    $6.8$  fixed              &                     \\
kT$_{\rm bb}$           &  $1.62 ^{+0.06} _{-0.06}$      &    $1.81  ^{+0.18} _{-0.15}$  &    $2.0  ^{+0.1} _{-0.1}$      &    $2.2 ^{+0.1} _{-0.1}$     &                      \\
R$_{\rm bb}$           &    $0.057 ^{+0.004} _{-0.003}$  &  $0.054 ^{+0.009} _{-0.007}$  &   $0.073  ^{+0.006} _{-0.005}$ &    $0.120  ^{+0.007} _{-0.007}$ &                               \\ 
$\chi^{2}_{\nu}$/dof    &   1.084/147                    &   0.630/26                    &    1.031/80                   &      0.941/105              &                    \\
\noalign {\smallskip}
\hline
\label{lstab:spec}
\end{tabular}
\end{center}
\end{table*}

	      \section{Discussion}\label{sec:discussion}

We reported here on the results of a 40 ks \xmm\ observation of the HMXB \src,
and on the re-analysis of $\cxo$ archival data.
The main result of our work is the discovery of QPOs in the mHz frequency regime in the \xmm\
observation, together with a periodicity in $\cxo$ data that could be ascribed to the spin period of the NS. 
Since there are several issues about \src\ that we would like to discuss,  we organize the discussion in different sub-sections, for clarity.

\subsection{Is \src\ associated with EXO~1912+097?}
\label{ls:skypos}

\begin{figure}
\begin{center}
\includegraphics[height=7.00cm,angle=0]{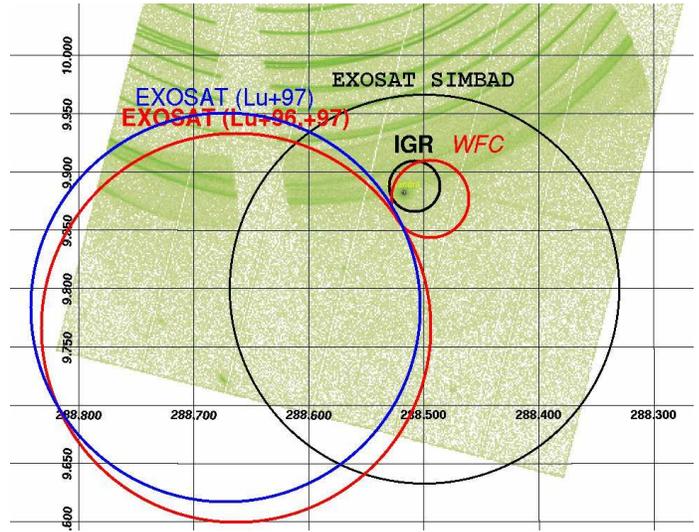}
\end{center}
\caption{\xmm\ EPIC pn image, together with the sky positions (and error circles as reported in Table~\ref{lstab:skypos}) of sources detected by different missions.
The $Chandra$ position coincides with the \xmm\ source. Equatorial coordinates are in units of degrees (J2000 equinox).
}
\label{lsfig:skypos}
\end{figure}

Several papers in literature  (\citealt{Hannikainen2004}, \citealt{Hannikainen2007}, \citealt{zand2004}, \citealt{zand2006}) proposed 
an association of \src\ with a poorly known EXOSAT source, EXO~1912+097. 
We show the sky region in Fig.~\ref{lsfig:skypos},  and report in Table~\ref{lstab:skypos} the coordinates of X--ray sources detected by different missions. We note that there is a typo in the WFC source coordinates reported in the text by \citealt{zand2004} (while the sky position is correctly shown in the figures by in't Zand et al. 2004, 2006). Therefore, we list in Table~\ref{lstab:skypos} also the correct WFC source coordinates (in't Zand 2016, private communication), for clarity.

Since in literature there is some discrepancy in the EXOSAT source sky coordinates
and its associated  uncertainty is also unclear, 
we summarize here the two original papers that reported this detection: \citet{Lu1996} and \citet{Lu1997} (this latter available only in Chinese).

\citet{Lu1996} analysed EXOSAT Medium Energy array (ME; 2-6 keV) scans of the Galactic plane using a direct demodulation method.
They listed only the source Galactic position, with no associated positional uncertainty (that we report in row n.4 of Table~\ref{lstab:skypos}).
An updated analysis of the same data, together with their statistical investigation, was reported in a later paper \citep{Lu1997}, where a {\em bootstrap} method
was used to derive the source centroid and its uncertainty. 
This source position from the original EXOSAT data reported by the 1997 paper is listed in Table~\ref{lstab:skypos} (row n.~5).
In row n.~6 we list the new centroid and its associated error radius derived from the {\em bootstrap} method \citep{Lu1997}.
From a visual inspection of the EXO~1912+097 map (source~1 in  Fig.~6 by \citealt{Lu1997}), 
we can derive an error radius of $\sim$10$\arcmin$ (at 95\% confidence level) in the source position.

Finally,  SIMBAD reports different celestial coordinates for EXO~1912+097 (as already noted by \citealt{zand2004, zand2006}).
This is puzzling, since the reference paper quoted by the SIMBAD website for the source sky coordinates is indeed Lu et al. (1996).
We report them in Table~\ref{lstab:skypos} (and in Fig.~\ref{lsfig:skypos}) 
as well, but with the caveat that they are likely wrong or, at least, of unknown origin.
Our feeling is that the SIMBAD source coordinates are simply rounded values from the EXOSAT name of the source.

To conclude, the $Chandra$ position  is only marginally consistent with the EXOSAT source reported by Lu and collaborators (1996, 1997),
being outside the EXOSAT 95\% error circle.

\begin{table*}
\caption{\label{lstab:skypos}
         X--ray sources detected by different X--ray missions, with their coordinates and position uncertainties. 
}
\begin{center}
\begin{tabular}{lllcc}
\hline
Observations                                                             &      R.A. (J2000)                      &    Dec. (J2000)                       &    Error (\& conf. level)      &       Ref. \\
\hline
Chandra                                                                  & $19^{\rm h}14^{\rm m}4\fs232$          &      $+9^\circ 52\arcmin58\farcs29$   &  $0\farcs6$          &   \citealt{zand2006}   \\
INTEGRAL                                                                 & $19^{\rm h}14^{\rm m}2\fs$             &      $+9^\circ 53.3\arcmin $          &  $1.3\arcmin$ (90\%) &   \citealt{Cabanac2004} \\
BeppoSAX/WFC                                                             & $19^{\rm h}13^{\rm m}58\fs8$           &      $+9^\circ 52\arcmin37\farcs0$    &  $2\arcmin$  (90\%)  &   in't Zand 2016 (priv. comm.)    \\
EXOSAT/ME  (EXO~1912+097; $l=44\fdg26$, $b=-0\fdg65$)                    & $19^{\rm h}14^{\rm m}39\fs2$           &      $+9^\circ 45\arcmin59\farcs4$    &      none            &   \citealt{Lu1996}      \\
EXOSAT/ME  (DD~1911+098; $l=44\fdg28$, $b=-0\fdg65$)                     & $19^{\rm h}14^{\rm m}41\fs47$          &      $+9^\circ 47\arcmin03\farcs2$    &    none              &   \citealt{Lu1997}      \\
EXOSAT/ME  (n.1 in Lu97 Table~5; $l=44\fdg26$, $b=-0\fdg65$)             & $19^{\rm h}14^{\rm m}39\fs2$           &      $+9^\circ 45\arcmin59\farcs4$    &  $10\arcmin$ (95\%)  &   \citealt{Lu1997}      \\
EXO~1912+097                                                             & $19^{\rm h}14^{\rm m}00\fs$            &      $+9^\circ 48\arcmin00\farcs0$    &      none            &   SIMBAD                \\
\hline
\end{tabular}
\end{center}
\end{table*}

\subsection{Did \xmm\ catch an X--ray eclipse?}

\xmm\  observed a very low level of X--ray emission from \src\ during the first $\sim$30~ks of the observation. 
This is the faintest state ever observed in this HMXB (2$\times$10$^{-12}$~erg~cm$^{-2}$~s$^{-1}$, 2-10 keV), translating
into a luminosity L$_{\rm X}$=3$\times$10$^{33}$~erg~s$^{-1}$ at a distance of 3.6 kpc \citep{Torrejon2010}.
This could rise the question of a possible eclipse of the X--ray source by the supergiant companion.
The rise in intensity approaching the end of the observation could suggest the eclipse  egress. 
Assuming the ephemeris MJD 51593.4$\pm{0.32}$ (epoch of maximum flux assuming a sinusoidal model) 
with the  period of 13.558 $\pm{0.004}$~days ($RXTE$/ASM; \citealt{Corbet2004}) 
the \xmm\ observation spanned an orbital phase interval $\Delta$$\phi$=0.51-0.54 
($\pm{0.12}$, extrapolating the uncertainties of 0.004~days on the orbital period to the present epoch). 
This implies that the \xmm\ observation falls near the minimum of the orbital light curve  ($\phi$=0 is the maximum in the light curve).
However, the orbital profile reported by \citet{Wen2006} ($RXTE$/ASM), does not show any evidence for an eclipse
during the X--ray minimum.
Moreover, the investigation of the variability of the intrinsic absorption along the orbit indicates a non-eclipsing system with a 
binary inclination around 65$^{\circ}$ \citep{Prat2008}.
Also our \xmm\ spectroscopy disfavours the possibility of an eclipse:
the X--ray spectrum of the very faint initial part of the observation does not show any of the 
typical properties of the scattered X-ray emission observed during an eclipse: 
an emerging iron line with a large equivalent width and a lower absorption 
compared to out-of-eclipse X-ray emission (because the scattering into the line of sight is by less dense wind material \citep{Haberl1991}). 
A prolonged absorption dip lasting the first half of the observation is also excluded, 
since the absorbing column density does not show evidence for variability along the whole observation.

Therefore, we conclude that the low intensity is intrinsic to the source, leading to 
at least 1000 the source dynamic range with respect to previous observations: \citet{Rodriguez2005} report on an X--ray flux,
corrected for the absorption, reaching 2$\times$10$^{-9}$~erg~cm$^{-2}$~s$^{-1}$ in the 1-20 keV energy range (see also \citealt{Rodriguez2006} for
another dim state, although 10 times brighter than what we observed with \xmm).

\subsection{\src: an intermediate case between SFXTs and persistent HMXBs?}

It is worth noting that a high dynamic range of 3 orders of magnitude in an HMXB is more typical of a Supergiant Fast X-ray Transient 
(SFXT; see \citealt{Sidoli2013} for a review) rather than of a persistent HMXB with a supergiant donor (SgHMXB).
In SFXTs the dynamic range is larger than 100 and it is one of the characterizing properties of the class, together
with the low duty cycle (a few \%) of their transient and luminous (L$>$a few 10$^{35}$~erg~s$^{-1}$) X-ray emission (see, e.g., \citealt{Paizis2014}).
\citet{Rodriguez2005}, from $RXTE$ and \inte\ observations, already pointed out a substantial variability 
of the source intensity.
We searched in the public \inte\ archive (12.3~yrs of data; \citealt{Paizis2013}) to get an estimate of the source duty cycle at hard X--rays: 
the source position falls within
the  \inte/IBIS FOV (within 12$^\circ$ from the centre) for about 9.68~Ms, and the source is detected for only  12\% of the time
in the hard band 17--50 keV.  
We can compare this duty cycle of 12\% (17--50 keV), with what observed with \inte\ from other kinds of SgHMXBs (all at similar distances, 2-4~kpc),
reported by \citet{Paizis2014}: the percentage of time spent in activity increases, starting from the most extreme SFXTs ($\sim$0.1\%), 
to the intermediate SFXTs ($\sim$3-5\%),   
to the transient SgHMXB 4U~1907+09 ($\sim$23\%), up to the persistent, eclipsing, SgHMXBs Vela~X-1 and 4U~1700-377 (76-79\%).
These latter two sources are basically always detected by \inte\ when they are not in eclipse.
Interestingly, the transient ``not so transient'' source 4U~1907+09 was suggested to be a sort of missing link between SFXTs and persistent SgHMXB 
 because of its intermediate level of activity \citep{Doroshenko2012}. 
This behaviour was confirmed at hard X-rays with \inte\  \citep{Paizis2014}.
In this respect, \src\ might be another example of a very variable SgHMXB, suggesting a smooth 
transition between SFXTs and persistent SgHMXBs.

\subsection{Quasi-periodic oscillations}
\label{sect:qpo}

We discovered QPOs in \src\ during the \xmm\ observation, 
with the strongest peak  at a frequency of 1.46$\pm{0.07}$~mHz, together with its second and third harmonics (Fig.~\ref{qpo}), 
in the second part of our observation,  at an X-ray luminosity $\sim$10$^{34}$~erg~s$^{-1}$.
Given its very low frequency, the total exposure of our observations limits 
prevented us from a further investigation of its evolution within the \xmm\ observation. 
However, its absence in $Chandra$ data indicates a transient nature.

QPOs in the range 1 mHz-20 Hz have been observed in 
several HMXB pulsars, with
persistent and transient X--ray emission \citep{Finger1998}, both 
with Be and early type supergiant companions (see \citet{James2010} for a list of sources). 
Usually these features are transient.
QPOs are also present in low mass X-ray binaries (LMXBs), but at much higher frequencies,
in the kHz range \citep{vanderklis1998}. 
The most popular models  explaining QPOs (both in LMXB and HMXBs) 
assume the existence of an accretion disc. 
While the Roche-lobe-overflow (RLO) in LMXBs guarantees the formation of a disc,
in HMXBs with no RLO its presence is much more elusive and debated: steady spin-up phases are usually 
taken as an evidence for it, while random walk spin behavior is believed to indicate
a wind-fed system. 
Spin-up phases are often observed in Be/XRBs systems during their
transient luminous outbursts, where an accretion disc is assumed to form from 
the dense slow wind coming from the Keplerian decretion disc of the massive Be companion. 
In HMXBs with supergiant companions with radially outflowing fast winds, the presence of a disc
is posed into question. Usually, both the short term variability of the X--ray flux 
and the pulsar spin period evolution, point to accretion from the companion wind.  
However, we will discuss QPOs in both kinds of accretion.

Typically, the QPO feature is interpreted as a diagnostic of the inner radius of 
the accretion disc feeding the NS and of its interaction with the
pulsar magnetosphere: it could be either the signature of
the Keplerian frequency of an accreting inhomogeneity moving at the magnetospheric radius, r$_{m}$ 
($\nu_{QPO}$=$\nu_{k}$(r$_{m}$); Keplerian frequency model, hereafter KM; \citealt{vanderklis1999}),
or its beat frequency with the NS spin  
($\nu_{QPO}$=$\nu_{k}$(r$_{m}$)-$\nu_{spin}$; 
beat frequency model, BFM; \citealt{Alpar1985}, \citealt{Lamb1985}).

If the periodicity we detected in $Chandra$ data is really the pulsar spin period,
in \src\ the frequency $\nu_{QPO}$ is much larger than $\nu_{spin}$, so both BFM and KM predict 
a similar Keplerian frequency, $\nu_{k}$, at the inner disc radius, r$_{in}$.
Since 
\be
r_{in} = \left({GM}\over{4 \pi^{2} \nu_{k}^{2}}\right)^{1/3}
\ee
assuming a NS mass, M, of  $1.4 M_{\odot}$, this implies r$_{in}$=1.2-1.3$\times$10$^{10}$~cm in \src.
In accreting X--ray pulsars it is possible to assume r$_{in}$$\equiv$r$_{m}$, where r$_{m}$ mainly depends 
on the NS dipole magnetic moment ($\mu$) and on the mass accretion rate $\dot M$ \citep{Ghosh1979}:

\be
r_{m}\equiv\eta\left(\frac{\mu^4}{GM\dot M^2}\right)^{1/7}
=(4.7\times10^9~{\rm cm})\,\eta\,\mu_{30}^{4/7}M_{1.4}^{-1/7}{\dot M}_{13}^{-2/7},
\label{eq:rm}
\ee
where $M=(1.4~M_{\odot})M_{1.4}$ is the NS mass,  $\eta$$\sim$0.5-1,
${\dot M}=(10^{13}\,{\rm g\,s^{-1}}){\dot M_{13}}$ is
the accretion rate, and $\mu_{30}=\mu/(10^{30}~{\rm G}\,{\rm cm}^3)$.
Therefore, in the BFM (and KM) framework, the above value for r$_{m}$ together with
the measured time-averaged X-ray luminosity
of L$_{2-10 keV}$=6$\times$10$^{33}$~erg~s$^{-1}$ during the \xmm\ observation (see Table~\ref{lstab:spec}), 
leads to a NS magnetic field strength B$\sim$(1-4)$\times$10$^{13}$~G.

To allow accretion, r$_{m}$ should be lower than the corotation radius 
(r$_{cor}$=($ GMP_{spin}^{2}/(4 \pi^{2}))^{1/3}$), the radial distance 
where the plasma in the disc corotates with the NS.
Under the already considered hypothesis that the 
spin period is really $P_{spin} = 5\,937\pm{219}$~s, we obtain a large corotation radius 
r$_{cor}$=5.4-5.6$\times$10$^{10}$~cm, which indeed enables accretion.
Since there is no independent constraint on the pulsar magnetic field,
we are not able to draw any unambiguous conclusion about the QPO interpretation.

However, given the supergiant donor (where a fast outflowing wind typically reaches a terminal velocity around 
1000~km~s$^{-1}$) and its wide orbit ($\sim$13.5 days), 
a wind-fed system is a more plausible picture for \src. 
In this case, a quasi-spherical settling accretion model for slow X--ray pulsars is applicable \citep{Shakura2012}.
In slow X--ray pulsars with a low X--ray luminosity (L$_{X}$$<$10$^{35}$~erg~s$^{-1}$) the wind matter captured within the Bondi radius of the NS
remains hot and cannot efficiently penetrate the NS magnetosphere. It accumulates above the magnetospheric radius, 
forming a quasi static hot shell of matter.
This extended quasi-static shell mediates both the accretion rate and the angular momentum transfer to (or removal from) 
the NS surface, by means of large-scale convective motions.
The characteristic time-scale for these motions produces the appearence of QPOs just at the mHz frequencies \citep{Shakura2012}, 
in agreement with what we observed in \xmm\ data. 

The fact that the \src\ QPO has a second and a third (rather than a fourth) harmonic
is quite curious (not easy to make odd harmonics), but this is not unheard of since it is also the case in
many BH LMXBs. Indeed the nature of QPOs' higher harmonics is an open problem in
BH binaries as well, where QPOs are very common features and are very well studied, especially at low 
frequencies.

It is worth noting that in all QPO models discussed above, the QPO frequency is expected to be  
correlated with the accretion rate. Assuming that the 
source flux is an appropriate proxy of the accretion rate, we expect to observe a higher QPO frequency at higher source fluxes. 
This seems to exclude 
an interpretation of the signal observed in the  \chandra\ data (0.17~mHz) as a QPO, because 
in this observation the flux was higher ($\sim$10$^{-10}$~erg~cm$^{-2}$~s$^{-1}$) 
than during the \xmm\ observation where the QPO frequency was 1.46~mHz. 
This favours the interpretation of the signal at $5\,937\pm{219}$~s as the NS spin period.

\section{Conclusions}

We discovered  1.46 mHz QPO (together with its second and third harmonics) 
from the HMXB \igr\  with \xmm, when the source
was in a faint state ($\sim$10$^{34}$~erg~s$^{-1}$).
Assuming the current models for the formation of QPOs in HMXBs (both in disc-fed and in wind-fed systems),
given the long orbital period and the supergiant nature of the donor star (implying a fast outflowing wind), 
quasi-spherical accretion from the companion wind is favoured. In this latter scenario,
the QPO is produced by the convective motions of material captured from the donor wind, 
that accumulates in a hot shell above the NS magnetosphere.
Our re-analysis of a public $Chandra$ observation shows a modulation at $5\,937\pm{219}$~s,
which we interpret as the spin period of the NS.
The comparison between the low flux shown during the \xmm\ observation and previous observations reported in literature, 
leads to about 3 orders of magnitude the dynamic range of the source intensity. 
It overlaps with the one observed from SFXTs, although the duty cycle of its luminous activity is not as extreme as in SFXTs. 
This suggests that \src\ is an intermediate system between them and the supergiant HMXBs with persistent X--ray emission.

\section*{Acknowledgments}

This work is based on data from observations with \xmm.
\xmm\ is an ESA science mission with instruments and
contributions directly funded by ESA Member States and the USA (NASA).
The scientific results reported in this article are also based on data obtained 
from the \cxo\ Data Archive. This research has made use of 
software provided by the \cxo\ X-ray Center (CXC) in the application package \textsc{ciao}, 
and of the \textsc{simbad} database, operated at CDS, Strasbourg, France.
This research has made use of the IGR Sources page maintained 
by J. Rodriguez (http://irfu.cea.fr/Sap/IGR-Sources/), and of the 
\inte~archive developed at INAF-IASF Milano, http://www.iasf-milano.inaf.it/$\sim$ada/GOLIA.html.
A.~Paizis and K.~Postnov are thanked for many interesting discussions.
LS is grateful to J. in't Zand  for help with the WFC source coordinates and for providing the text of the Chinese paper Lu et al (1997).
LS thanks Silvia Sidoli for the translation of the Chinese paper Lu et al (1997) and 
acknowledges the grant from PRIN-INAF 2014,  ``Towards a unified picture of accretion in High Mass X-Ray Binaries''  (PI: L.~Sidoli).
PE acknowledges funding in the framework of the NWO Vidi award A.2320.0076 (PI: N.~Rea).
SEM acknowledges the Glasstone research fellowship program.
\bibliographystyle{mn2e} 
\bibliographystyle{mnras}

\bsp

\label{lastpage}

\end{document}